# State-Based Random Access: A Cross-Layer Approach

Amir M. Khodaian, Babak H. Khalaj
Department of Electrical Engineering, and Advanced communication Research Institute
Sharif University of Technology, Tehran, Iran
Email: khodaian@ee.shrif.edu, khalaj@sharif.edu

*Abstract*— **In this paper, we propose novel state-based algorithms which dynamically control the random access network based on its current state such as channel states of wireless links and backlog states of the queues. After formulating the problem, corresponding algorithms with diverse control functions are proposed. Consequently, it will be shown that the proposed state-based schemes for control of the random access networks, results in significant performance gains in comparison with previously proposed control algorithms. In order to select an appropriate control function, performances of the state-based control algorithms are compared for a wide range of traffic scenarios. It is also shown that even an approximate knowledge of network statistics helps in selecting the proper state dependent control function.**

## I. INTRODUCTION

In wireless communication networks a variety of performance criteria such as rate, energy, and delay are prioritized based on a specific application and limitations of the network and its elements. These criteria are controlled in different parts of a network, such as rate control at sources or scheduling at network links. Most of communication network protocols investigate these problems and propose algorithms to control network parameters.

Optimal control of network parameters has been the subject of many researches. Optimal flow control was first introduced in a seminal work by Kelly *et. al.* [1] where proportional fairness is achieved based on the proposed shadow prices for transmission on each link. This subject was extended to cross-layer optimization in other works [2]-[6] and optimal parameters were obtained using mathematical programming. In order to realize a practical algorithm, the optimization problem was decomposed between network layers and nodes in order to achieve distributed solutions. Most of the aforementioned work assumed static parameters that did not change over time and try to find optimal values for such parameters. However, due to inherent network dynamics, dynamic control of a network may in general result in higher performance; therefore an optimal algorithm should continuously control network parameters. This line of research is investigated in more detail in optimal scheduling problems.

In a pioneer work, Tassiulas and Ephremides [7] proved maximum weight (MW) scheduling, which schedules links based on the backlog state of the queues in order to stabilize the largest throughput region. However, the proposed algorithm is impractical due to its high complexity and its need for a centralized controller. In order to achieve a practical scheduler, different algorithms are proposed. Among them, the queue-based control of random access networks has recently attracted much attention [8], [9]. Recently, it is also shown that in single-hop random access Carrier Sense Multiple Access (CSMA) networks, there exists low complexity distributed algorithms that can achieve the maximum stable throughput region [10].

Our paper takes the first step to show that optimal control of random access network should be generally based on the *network state* in different layers and different network elements. Previously, Tong, *et.al.* [11] utilized Channel State Information (CSI) to control the random access probability. The idea was later extended to multi-hop networks in [12] where multiuser diversity was exploited. However, none of the aforementioned papers considered queue state which adversely affects delay performance of the proposed algorithms. Queue-based random access was studied in [8]-[10] where stability of the simple proposed algorithms were verified and their delay performance were compared with static algorithms. However, in such studies different queue-based algorithms were not compared with each other. In addition, CSI has not been considered in control algorithms used in such schemes.

The rest of the paper is organized as follows. The network model is presented in the next section. Then, in section III we formulate and investigate the problem. We propose possible suboptimal solutions in section IV. Section V explores the provided state dependent algorithms via numerical results and comparisons. Finally, we conclude the paper and discuss possible extensions in section VI.

## II. NETWORK MODEL

Suppose an ad-hoc wireless network consisting of *N* nodes that use the set of links *L* to transmit packets from source nodes to destination nodes. We assume existence of a predetermined path for each source and destination pair (session). Each node and its outgoing links have a separate queue for the sessions that pass through them. Number of packets waiting for service in queue of link *l* which belong to session *s*, $q^s_l(t)$, forms state of the link queue at slot *t*. Nodes are assumed to have infinite buffers, thus, there is no packet drop in the network and the

This work is supported in part by Advanced Communication Research Institute (ACRI) and Iran Telecommunication Research Center (ITRC).

network will be unstable if backlog in any of the queues go to infinity.

Link $l$, with a non-empty queue, will transmit packets that belong to session $s$ with probability $p_l^s(t)$ using the slotted Aloha protocol using the same slot time and length throughout the network. In the proposed algorithms, this probability will be controlled dynamically at each slot based on the network state. $P_i(t)$ is the transmission probability of the node $i$, which is equal to the sum of transmission probabilities of its output links for all sessions.

Nodes transmit synchronously in the network and time slot duration ($T_s$) is a multiple of the minimum packet transmission time $\theta$. We assume all sources in the network generate packets with the same size and a link can transmit an integer number of these packets at a time slot using channel-state dependent adaptive modulation. $c_l(t)$ denotes the number of packets that link $l$ can transmit at time slot $t$. Collision happens if one of its neighbors start transmission while receiving a packet. The transmission or service rate of link $l$ for session $s$, denoted by $x_l^s(t)$ depends both on successful packet transmission (without collision) and the channel state at the transmission slot.

The acronyms used for link and node sets are described in Table I.

TABLE I.  ACRONYMS

| Acronym | Definition |
|---|---|
| $N_i$ | Set of neighbors of node $i$. |
| $S$ | All active network sessions. |
| $S_l$ | Set of sessions that visit link $l$ along their path. |
| $L_s$ | Set of links belonging to path of session $s$ |
| $L_l^I$ | The set of links that transmission on link $l$ will interfere with them. |

In this paper, we assume all nodes have equal transmission power, resulting in symmetric neighborhoods.

### III. PROBLEM FORMULATION

Performance of an ad-hoc network depends on variety of parameters with the most important ones being source rate, transmission delay, energy consumption, and fairness among similar nodes, where each of these performance criteria can be less or more important based on the application. For example in a sensor network, energy consumption is a critical parameter while in mesh networking throughput and delay are more important. Also, in some applications a threshold (constraint) may be set for the provided bandwidth or delay while others are only looking for more bandwidth and less delay. In this section we will provide the mathematical formulation for the problem and investigate some special cases in more details. General formulation of such problems is provided in (1).

$$\begin{aligned} \min \quad & G(x_s, d_s, E_s) \\ s.t. \quad & \left.\begin{array}{l} d_s < D_s^{th} \\ x_s < X_s^{th} \end{array}\right\} \forall s \in S \\ & \text{Capacity constraints} \end{aligned} \quad (1)$$

where $x_s$, $d_s$, and $E_s$ represent short or long term average of rate, delay, and energy for packets that are transmitted over session $s$. The cost function $G(\cdot,\cdot,\cdot)$, prioritizes sessions based on their requirements, provides fairness among nodes, and controls tradeoff among rate, delay, and energy consumption in the network.

For a stable network without considerable variations in the dynamics, we can use $\bar{x}_s$, and $\bar{d}_s$ as the long term average of the rate and delay seen by session $s$. For example, in [6] a special case of (1) with average rate and energy in the utility function and a constraint over average end-to-end delay, is solved for random access networks. The solution provided finds optimal transmission probabilities assuming a static framework. However for stochastic networks with considerable variations, short-term performance of the network is also important and should be considered in design.

We define the performance evaluation duration $T_E$ which is a set of timeslots that will be used to consider the effect of the current decision. $T_E$ which contains current time $t$ may include time slots in past and future. For example, (2) is the transmission rate of session $s$, which is defined over $T_E$ in order to illuminate the effect of current transmission rate on rate performance of the network throughout the whole $T_E$ period:

$$x_s^{T_E} = \sum_{t' \in T_E} x_s(t') \quad (2)$$

Furthermore, link capacity constraints correspond to stability of the link queues: $q_l^s(t) < \infty, \forall t, \forall l$. Considering the Little's law, we may also use summation of the link queue lengths as an approximate measure for end-to-end or session delay. The end-to-end delay constraints of (3) also include stability of links and, therefore, satisfy capacity constraints.

$$\left( \sum_{t' \in T_E} \sum_{l \in L_s} q_l^s(t') \right) < D_s^{th} x_s^{T_E}(t) \quad (3)$$

Energy consumption of the network's links over $T_E$ can be defined as:

$$E^{T_E}(t) = \sum_{t' \in T_E} \sum_{l \in L} E_l(t') \quad (4)$$

A similar approach is used in [13] where delivery contract is defined as a constraint over total flow during a given time interval. However, energy consumption and queuing dynamics are not considered and no constraint is set for the end-to-end delay of the flows.

In order to formulate the optimization problem, first we suppose $G(x_s, d_s, E)$ to be a linear cost function with the coefficients $\alpha_s$, $\beta_s$, and $\delta$ for session rates, end-to-end backlogs and energy consumption, respectively. The objective function can be written as:

$$\min \sum_{t' \in T_E} \left\{ \sum_{s \in S} (-\alpha_s x_s(t') + \beta_s \sum_{l \in L_s} q_l^s(t')) + \delta \sum_{l \in L} E_l(t') \right\} \quad (5)$$

It is easy to observe that (5) has the formulation of a Dynamic Programming (DP) problem [14]. Time slots are stages of the dynamic system, queue backlogs $q_l^s(t)$ form system stated, and link transmission probabilities are either control or decision variables. The system function will depend on channel states which are independent random variables:

$$q_l^s(t+1) = \begin{cases} q_l^s(t) - \rho_l^s(t)c_l(t) + x_s(t) & \text{at source} \\ \begin{pmatrix} q_l^s(t) - \rho_l^s(t)c_l(t) + \\ \rho_k^s(t)c_k(t) \end{pmatrix} & \text{other links} \end{cases} \quad (6)$$

where $\rho_l^s(t)$ is the probability of transmission without collision over link $l$ at slot $t$, and $k$ is the link preceding link $l$ in session $s$.

In order to solve problem formulated in (5) and provide optimal transmission probabilities and source rates for all time slots, the information about current and future system states is required. Also the optimal values depend on future decisions and typical solutions to such dynamic optimization problems will approximate the future states and decisions. These problems are too complex and possible distributed solutions require large message passing.

In addition, for non-linear cost functions, which are required to consider fairness, the problem is not a basic dynamic optimization problem. For example, the rate utility function of the network which is shown in (7) cannot be easily decomposed over time.

$$x_s^{T_E}(t) = \sum_{s \in S} \log \left( \sum_{t' \in T_E} x_s(t') \right) \quad (7)$$

In the following, we will formulate and solve some special cases of the problem for selected networks. Subsequently, in section IV, heuristic algorithms will be provided that can be applied to general random access networks.

*A. Single Link*

Consider the case of a single source-destination pair as a simple example. We consider the objective function at $t$ to be a nonlinear combination of transmitted packets and consumed energy during the interval $T_E = \{t - N : t\}$:

$$\log \left( \sum_{t' \in T_E} p(t')c(t') \right) - e\delta \sum_{t' \in T_E} p(t') \quad (8)$$

In order to maximize (8) at $t$, we use variable $X(t)$ to denote the amount of the transmitted packets prior to $t$:

$$X(t) = \sum_{t' \in \{t-N : t-1\}} p(t')c(t') \quad (9)$$

Then, optimal link transmission probability can be calculated as:

$$p^*(t) = \text{proj}_{[0,1]} \left( \frac{1}{e\delta} - \frac{X(t)}{c(t)} \right) \quad (10)$$

where $\text{proj}_{[0,1]}(x) = \min(\max(x, 0), 1)$. Also note that $p^*(t) = 0$ when there is no packet available in the link queue.

Thus, optimal transmission probability is proportional to channel state and is inversely proportional to packet transmission energy and amount of previously transmitted packets.

*B. Single-hop Network*

Consider a network with $n$ source-destination pairs. We set the objective function to be a linear combination of rate utility and energy consumption and assume no constraint over rate or delay. Consequently, the optimization problem will reduce to:

$$\max \sum_{l \in L} \log \left( \sum_{t' \in T_E} p_l(t')c_l(t')(1 - \sigma_l(t')) \right) - e\delta \sum_{l \in L} \sum_{t' \in T_E} p_l(t') \quad (11)$$

Session indexes are omitted since there is a unique session per link in the single hop network. We assume that nodes can estimate collision probability, $\sigma_l(t)$ for transmitted packets. It is assumed that each link will use the same energy for packet transmission.

Similar to the definition in (9), we define $X_l(t)$ as the amount of successfully transmitted packets over link $l$. Also, $Y_l(t)$ is defined as the amount of energy consumed for transmission over link $l$ prior to $t$. Consequently, (11) can be rewritten as:

$$\max \sum_{l \in L} \log \left( p_l(t)c_l(t)(1 - \sigma_l(t)) + X_l(t) \right) - \delta \left( ep_l(t) + Y_l(t) \right) \quad (12)$$

We can then calculate optimal transmission probability on link $l$ as:

$$p_l^*(t) = \text{proj}_{[0,1]} \left( \frac{1}{e\delta} - \frac{X_l(t)}{c_l(t)(1 - \sigma_l(t))} \right) \quad (13)$$

Therefore, transmission probability should be decreased as collision probability increases. Collision probability can be estimated using the statistics of the transmitted packets on channel at earlier timeslots and, also, the queue state of neighbors.

IV. STATE-BASED ALGORITHMS

Due to complexities described for optimal solution of the general random access network in the previous section, in this section we will provide suboptimal algorithms that will control the random access network based on the network state. These algorithms should define functions that will select $p_l^s$ and $x_s$ at each time slot.

It is known that adaptation of the parameters to the network variations is vital. Before presenting the suboptimal state-based algorithms, let's have a quick look at the weighted proportional fairness problem in random access:

$$\max \sum_{l \in L} w_l \log(r_l) \quad (14)$$

where $r_l$ is the transmission rate of the link $l$ which should be less than or equal to its throughput and $w_l$ is the weight assigned to the link $l$. This problem is solved in [15] and a non-iterative distributed algorithm is provided which uses (15) to calculate link transmission probabilities:

$$p_l = \frac{w_l}{\sum_{k \in L_l^l} w_k} \quad (15)$$

where, $L_l^I$ is defined in 0 If we set link weights to depend on the queue and channel states, this non-iterative solution can be utilized to provide transmission probabilities based on the current network state.

### A. Channel and queue states in the weight function

One important question is how channel and queue states should affect link weights. Optimal scheduling for variable channel state was considered in [16] and it was shown that the max weight problem, (16), should be solved in each slot throughout the network in order to get the maximum stable rate region.

$$\max_{I(t)\in I}\{\sum_{s\in S}\sum_{l\in L}\Delta q_l^s(t)x_l^s(t)\} \quad (16)$$

where, $I(t)$ is the network control policy that will determine decisions made throughout the network, and $\Delta q_l^s(t)$ is the differential queue backlog between link $l$ and the next link of session s.

In random access networks $x_l^s(t)$ depends on channel state $c_l(t)$ and successful transmission probability at slot $t$, $\rho_l^s(t)$. As a result, the problem (16) can be viewed as a weighted maximization problem where the weight of each link is the multiple of its differential queue backlog and its channel state:

$$w_l^s(t) = \Delta q_l^s(t)c_l(t) \quad (17)$$

### B. Transmission probability control functions

The core of random access control is setting the appropriate transmission probabilities for links. For example, the binary exponential back-off algorithm in IEEE 802.11 standard adapts transmission probability at each slot based on the collision or successful transmission at earlier slots.

Here, we will use the link weights defined in section IV.A for random access control. Considering the weighted proportional fairness results in (15), we should set the transmission probability of link $l$ based on its weight and weight of the links that transmission on $l$ can interfere with them. Thus, we can introduce the primary proposed function:

$$p_l^s(t) = \frac{w_l^s(t)}{\sum_{s'\in S}\sum_{k\in L_l^I}w_k^{s'}(t)} \quad (18)$$

We call the control algorithm that uses (18) by the acronym linear-SBRA (State-Based Random Access). We also introduce other functions of link weights in numerator and denominator. We may use square of the link weights to obtain square-SBRA:

$$p_l^s(t) = \frac{(w_l^s(t))^2}{\sum_{s'\in S}\sum_{k\in L_l^I}(w_k^{s'}(t))^2} \quad (19)$$

or the exponential function (exponential-SBRA):

$$p_l^s(t) = \frac{\exp(w_l^s(t))}{\sum_{s'\in S}\sum_{k\in L_l^I}\exp(w_k^{s'}(t))} \quad (20)$$

We will show in the numerical results of section V that proper selection of the transmission probability function depends on the statistics of the arrival traffic rates at network sources.

## V. NUMERICAL EXAMPLES

In this section, we provide the results of applying the proposed algorithms. Our goal is to discuss the behavior of proposed algorithms and compare them in the case of small networks.

### A. Single link

Consider the case of a single link and select the transmission probability on this link using (10). We assume arrival at the link to be Poisson and transmission on this link during a timeslot consumes unit energy, $e = 1$. The performance evaluation period is also assumed to be of length $N = 10$. We assume a timeslot duration of 5msec and $i.i.d$ rounded Rayleigh channel at each slot with average channel rate of 1. The results of simulating the algorithm for different arrival rates, $r$, and using two values for the energy coefficient $\delta$, $\delta_1 = \frac{1}{N\times r}$ and $\delta_2 = \frac{\delta_1}{10}$, is shown in Table II.

TABLE II. SIGNLE LINK

| Arrival | Delay, $\delta_1$ | Delay, $\delta_2$ | Energy, $\delta_1$ | Energy, $\delta_2$ |
|---|---|---|---|---|
| 0.3 | 0.02 | 0.0004 | 0.23 | 0.29 |
| 0.5 | 0.045 | 0.0008 | 0.36 | 0.48 |
| 0.7 | 1.6 | 0.001 | 0.048 | 0.65 |
| 0.9 | 3.8 | 0.006 | 0.61 | 0.79 |
| 1 | 4.6 | 0.111 | 0.67 | 0.82 |

It is evident that the tradeoff between delay and energy can be controlled with $\delta$. However, simulation results show that delay will change significantly with little decrease in energy consumption.

### B. Multi-hop Adhoc network

As a sample of an ad-hoc network, we have selected the network of Fig. 1 with 10 nodes, 12 links and 4 sessions. It is assumed that there is a separate queue for each session on the links. Similar assumptions as section V.A are used for channel and links. The arrivals at session sources are assumed Poisson with the same rate.

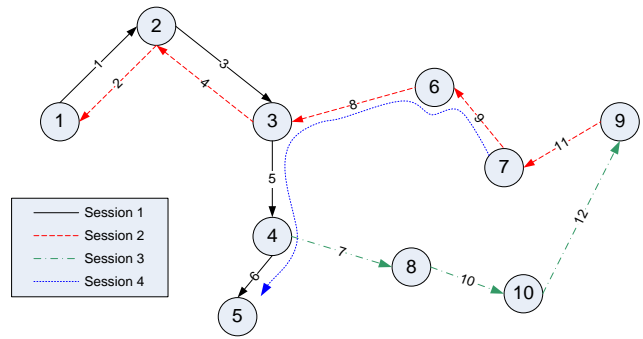

Figure 1. Sample network

Performance of state-based random access algorithms used over this network is compared with each other and with the queue-based algorithm proposed in [9]. We have assumed that the channel state updates every 10 slots, also link transmission probabilities will be updated every 3 slots. Delay performance of different algorithms is compared in Fig. 2. As shown in this figure, the 1st order SBRA algorithm has the best performance for the low-traffic region. It is apparent that exponential SBRA can tolerate the highest rates.

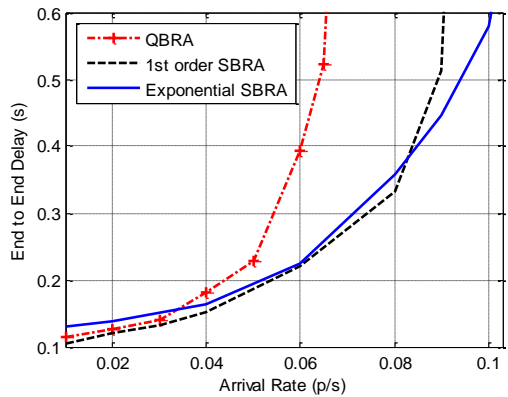

Figure 2. Comparison of queue- and state-based algorithms

Thus, the appropriate algorithm should be selected based on the arrival rate at the network.

## VI. CONCLUSION AND FUTURE WORKS

In this paper, we have formulated the optimal random access control problem with general objective functions and constraints. It is shown that the special cases of the optimal random access network problem can be formulated as a dynamic programming. The problem is solved for single link and single hop networks. The optimal transmission probability of a link is shown to be proportional to channel state and inversely proportional to collision probability and amount of previously transmitted packets. We have also provided some heuristic state-based algorithms and showed that selecting the best algorithm depends on the knowledge of the arrival statistics.